# A Study of Solar Irradiance Prediction Error Impact on a Home Energy Management System

Mallek Sallami Mziou
IRT SystemX
Palaiseau, France
e-mail: mallek.mziou-sallami@irt-systemx.fr

Rim Kaddah
IRT SystemX
Palaiseau, France
e-mail: rim.kaddah@irt-systemx.fr

Amira Ben Hamida
IRT SystemX
Palaiseau, France
e-mail: amira.benhamida@irt-systemx.fr

*Abstract*— Nowadays, Energy Management Systems (EMS) are accessible for homes and buildings to optimize energy consumption especially when solar panels and batteries are installed. The intelligence of existing systems is often based on environmental or exogenous information like the weather, energy prices, and endogenous information like user consumption behavior and activity. The solutions aim to adapt a consumption profile to the produced energy in order to reduce costs. In the case of a perfect prediction of all variables, system performance can be controlled. In this article, we study the impact of generation prediction error on the daily energy cost. For this, we consider the energy management system as a black-box and we simulate multiple scenarios with different prediction errors using the quasi-random Monte Carlo method. We observe the global sensitivity of the system by measuring the Sobol indices in order to identify errors that impact more the daily energy cost. The analyses are based on French consumption data and on irradiance data for Carpentras, France. Results show that findings are aligned with battery charge and discharge strategies.

## I. INTRODUCTION

With the current environmental issues, several solar energy-based technologies have been proposed as a partial solution to reach environmental goals. In the context of smart homes, the benefit of using green energy can be both environmental and economic. However, due to the solar resource high variability and depending on the energy purchase and selling schemes, forecasting algorithms can be important to better plan with respect to incoming solar production. Using forecasting information, flexible energy devices functioning can be optimized with respect to local generation and by taking into consideration the wear of the equipment. So, a measure of performance is required. Indeed, a better understanding of uncertainty in the netload (consumption – generation) allows to better maintain grid stability and develop adequate real-time control mechanisms. This can have the positive effect of keeping down the costs.

Furthermore, the topic of coping with uncertainty for complex systems is not new [1]. In addition, researchers have developed systems to simulate buildings thermal and energy behavior as black-box functions including many parameters. Due to the complexity of buildings energy model, only few important parameters are taken into account.

In this paper, we study the impact of forecasting errors on the total daily system cost. For this, we analyze the sensitivity of irradiance prediction errors on this total cost.

To perform a Sensitivity Analysis (SA), several approaches and categorization were proposed in the literature. According to [2], SA methods may be classified into three approaches: the mathematical approach, the statistical approach and graphical assessment. The major difference between these methods relates to the number of input parameters. For instance, in the statistical sensitivity analysis, a large enough number of inputs has to be (randomly) generated.

Heiselberg et al. [3] proposed to group Sensitivity Analysis methods into three classes, which are: the local sensitivity methods, the global sensitivity methods, and the screening methods. In the local sensitivity methods, one can study the variation of system output under the variation of one parameter. In global methods, the sensitivity to one input is computed by varying simultaneously many other inputs (i.e., input is represented by a vector). Screening methods compute the sensitivity indices as an average of derivative with respect to the different inputs. In the context of the building energy models, the choice of the most suitable SA method depends on the assumptions one can make on the model output and mainly on the linearity of the function linking inputs to the output.

A similar SA analysis has been conducted by authors in [4]. The authors present a performance comparison of sensitivity analysis methods for building energy model in terms of time and computation. However, the interpretability of the sensitivity indices is lacking. In addition, their analysis is solely related to physical endogenous variables of the house and does not take into account exogenous variables that may affect energy management system decisions.

In this work, we focus on a global SA for the assessment of the Energy-Efficient Smart Home solutions. To conduct this analysis, we use a hybrid modular simulator separating the user input, optimization module and physical simulation module. The cited components are connected thanks to a loosely-coupled architecture that enables communication exchanges.

For the SA analysis, we use the Sobol method [5] [6] for evaluating the system's behavior by performing a functional analysis of the variance (ANOVA). To generate the scenarios to be evaluated, many strategies were proposed like Sobol sequence [6], Fast method [7], etc. It is also possible to generate a sequence of randomly distributed points of inputs using the Monte Carlo method. In practice, it is common to substitute random sequences with low discrepancy sequences to improve the efficiency of the estimators. This method is known as the quasi-Monte Carlo method. It is less expensive than Monte Carlo, but no evaluation of the error is made. The third one is quasi-Monte Carlo randomized (intermediate cost between the two previous methods and evaluation of the error). In this study, we use the Saltelli's sampling scheme, which is based on quasi-random Monte Carlo [12].

The remainder of this paper is structured as follows: in Section II, we describe our system: the software system architecture and simulation modules. We also describe the home energy model with a specific highlight on the optimization and the sensitivity analysis procedure considered. In Section III, we present the sensitivity study results and analyze the correlation between the daily total cost and the forecast error. Finally, we conclude in Section IV.

## II. SYSTEM

We aim at building a system capable of analyzing the impact of prediction errors on an energy management system. In particular, we are interested in studying the impact of irradiance prediction errors on household electricity bills. We focus on a residential house equipped with solar panels and a storage device (see Figure 1. ). This house can have varying buying and selling prices.

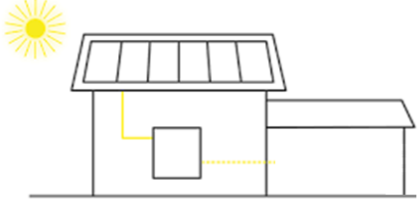

Figure 1.  Smart Home with solar panels and a storage device

### A. Smart Home architecture

To assess the impact on a realistic house, we build a user-friendly Smart Home simulator. It is designed as a modular platform: each component of the platform represents a different function and can be replaced and/or developed independently. It contains three main components:

1. *The User Interface (UI) component:* allows the user to enter all the required inputs and initiate the simulation. It helps to select the equipment of the Smart Home. Finally, the UI renders the results of the study in a summarized report.
2. *The Optimization component*: calculates and finds the best planning for controllable devices in the Smart Home taking into account the constraints of electric equipment, buying and selling price and local generation of renewables.
3. *The Simulation component:* simulates the physical response of the house taking into account the equipment used and the planning generated by the optimization component. Our model is built using Phisim library [8].

To ensure the communication, we implement messaging between each of the previously mentioned components. For this, we use the Websocket technology and nodejs language. In Figure 2. , we present the platform including all the components. Grey boxes represent different distant machines. In our analysis, the user is an automated script that launches scenarios corresponding to the conducted sensitivity analysis.

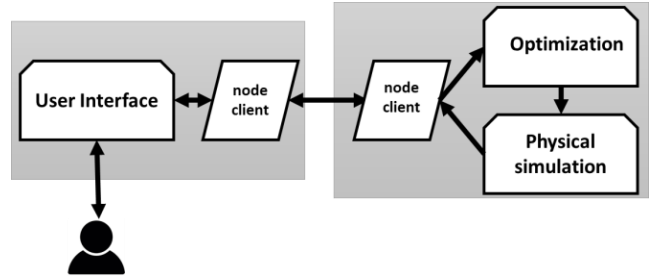

Figure 2.  Smart Home simulator structure

### B. Context and assumptions

We consider an energy management system that works as follows: each day, the system produces a charge and discharge planning for each hour of the next day. This case allows to decide energy trading decisions on day-ahead market coupled with having the possibility to sell at a feed-in tariffs. In this paper, we do not consider intra-day control mechanisms but we aim at understanding the uncertainty that can occur when we are planning ahead of time.

The energy management system decides to auto-consume or sell the energy produced by the solar panels. This is done based on a house consumption and energy production forecast. The system does not change the house consumption habits (no demand management is considered for our analysis). In addition, the use of battery is restricted to fulfilling house consumption needs (i.e., the energy stored in the battery cannot be sold to the grid).

### C. Optimization model

Based on solar irradiance forecasts, the system solves a cost minimization problem formalized as a Mixed Integer Linear Problem (MILP):

$$\min \sum_{h=1..24} (Q_h P_{buy}(h) - p_h P(h) P_{sell}(h)) \quad (1a)$$

Such that:

$$Q_h = Y(h) + B_h - (1 - p_h) P(h) \quad (1b)$$
$$p_h \in [0, 1] \quad (1c)$$
$$Q_h \geq 0 \quad (1d)$$
$$B_h \in [-B_{maxDischarge}, B_{maxCharge}] \quad (1e)$$

where h is an hour of a day (h=1,..,24), $Q_h$ is the amount of energy bought at hour h at the price $P_{buy}(h)$, $p_h$ is the fraction of the power produced P(h) by solar panels at hour h. This power can be sold at the price $P_{sell}(h)$. Y(h) denotes the consumption forecast at hour h. The battery charge or discharge rate at hour h is denoted by $B_h$. It takes values between a maximum discharge rate $-B_{maxDischarge}$ and a maximum charge rate $B_{maxCharge}$. The evolution of the energy content of the battery $E_h$ at hour h is given by:

$$E_h = \begin{cases} E_{h-1} + \eta B_h & \text{if } B_h \geq 0 \\ E_{h-1} + B_h & \text{Otherwise} \end{cases} \quad (2)$$

This energy content is a positive value smaller than the considered battery capacity. η represents the battery charging efficiency.

The optimization system is linear. To make a realistic assessment of the house response, we run the optimized planning in a physical simulator, which tries to imitate the real energetic behavior using Phisim library [8].

*D. Sensitivity analysis model*

There are many approaches to perform sensitivity analysis. Since we do not want any restriction on the energy management system model, we use a black-box method consisting in running simulations and observing the effect on the output.

In addition to the nature of the analysis technique, the choice of the method also depends on the variables dependencies. We can study the impact of variables separately. This is called a local sensitivity analysis. Or, we can analyze the impact by varying variables simultaneously. This is called a global sensitivity analysis. In our case, since we are interested in analyzing the impact of solar generation on costs, variables are naturally related. Indeed, we can expect three main causes of differences between predicted and real irradiance, namely a shift with respect to prediction, a change in the amplitude on the whole irradiance curve or unexpected local (hourly) changes (e.g., caused by clouds). For this reason, we argue that a global analysis method is more appropriate.

The global sensitivity is often measured by a numeric value called the global sensitivity index. This index can be of three orders (see Figure 3. for illustration):
- **First-order index:** measures the contribution of a single model input (alone) to the output variance.
- **Second-order index:** measures the contribution of the interaction of two model inputs to the output variance.
- **Total-order index:** measures the contribution to the output variance of a model input, including both its first-order effects (the input varying alone) and all higher-order interactions.

Suppose a model denoted by *G* and its *d* input variables denoted by $X_1, X_2, …, X_d$. The output Y can be described by:

$$Y = G(X_1, X_2, …, X_d) \quad (3)$$

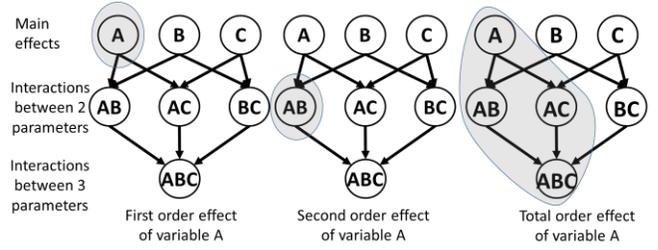

Figure 3. Different orders of sensitivity indices

To estimate the sensitivity indices, several methods exist in the literature. We compute sensitivity indices obtained through a variance decomposition of G (see [9] for details). They satisfy:

$$1 = \sum_{i=1..d} SI_i + \sum_{i<j} SI_{i,j} + … + SI_{1,2,…,d} \quad (4)$$

where $SI_i$ is the principal (First order) sensitivity index of variable $X_i$, $SI_{i,j}$ is the second order index of variables $X_i$ and $X_j$. The total sensitivity indices are defined based on the first order indices with these equations:

$$SI_i = Var(E(Y/X_i))/Var(Y) \quad (5)$$
$$SI^T_i = \sum_{I \subset \{1,…,d\}, i \in I} SI_I \quad (6)$$

where $SI^T_i$ measures the contribution of $X_i$ to the output variance including the variance caused by its interactions with the other input variables. Estimating these indices can be costly in time and computation [10] [11]. In this paper, we use the method described in [12]. It is based on the Saltelli's sampling scheme, which is based on quasi-random Monte Carlo.

### III. APPLICATION

In this section, we propose to conduct the sensitivity analysis on the proposed home energy management system. In our model, the intelligence of the smart home management relies on the optimization results. The performance of these results depends on the accuracy of solar irradiation prediction. However, there is always a difference between the predicted and the real irradiation.

*A. Objectives*

The aim of this work is to take advantage of the knowledge of the energy management system strategy in order to assess the interpretability of the sensitivity analysis results. For this, we assume that prediction error can occur at any of the 24 hours and we measure the impact of the prediction error on the total cost at the end of the day. Since, the optimization controls the battery charge and discharge, we include the electric storage capacity as an additional parameter that may impact sensitivity. Therefore, different sizes of the battery and prediction error have to be studied.

*B. Model simplification*

Simulator inputs can be grouped into three categories:
- Installed equipment in the house, with their characteristics and limitations (electric power, flexibility, etc.).

- The forecasted environmental and market variables: temperature, irradiance, wind speed, electricity buying prices, electricity selling prices.
- The occupation scenario of the different equipment based on user activity, or an expected behavior of consumption devices.

Conducting sensitivity analysis of all these variables can be very costly in time and is not in the scope of the present paper. In order to reduce the complexity of our system, we fix the parameters representing hourly electricity prices and energy consumption profile.

To select our model input for the house consumption, we use the results of the analysis in [13], which identifies four typical consumption profiles for houses in France represented by four periods: summer week, summer weekend, winter week and winter weekend. This is built based on consumption data collected from 149 houses and 36 multiple-unit dwellings of different characteristics in 2010. Since we are interested in studying solar generation, we consider the summer profile of a weekday.

To ease the interpretation of the sensitivity analysis results, we use simple pricing models for energy purchase and selling. Indeed, we consider a fixed buying price at 0.2977€/kWh and a selling price at 0.1231€/kWh. This pricing scheme is very similar to the actual German pricing for feed-in-tariff and energy purchase (selling price lower than the buying price). It incentivizes to auto-consume PhotoVoltaic (PV) power. Average solar irradiance is calculated using data from Carpentras station (France) for 2011 to 2013 during summer. We use average generation values of a solar panel with a performance ratio of 15%.

The fixed daily scenario considered with respect to energy price, consumption and generation profiles, is represented in Figure 4.

We consider four classes for battery capacity (kWh):
- First class: ranges from 5 to 8,
- Second class: ranges from 9 to 12,
- Third class: ranges from 15 to 20,
- Fourth class: ranges from 30 to 40.

First and second class correspond to battery capacities that are lower than 12933Wh (the total consumption after 4pm that cannot be satisfied by the mean predicted PV generation). Third and fourth classes correspond to batteries larger than the needed power storage.

In this study, we use the Saltelli's sampling scheme to generate the scenarios for irradiance variables and battery capacity. In order to validate our observations, we conduct the study also by shifting the generation curve to the left (i.e., the generation peak is earlier) and to the right (i.e., the generation peak is later).

*C. Results and interpretations*

To analyze the sensitivity of the model with respect to irradiance and battery capacity, a N(2d+2) sample matrix is generated where d is the number of parameters (related with hours and capacity) and N is a number of scenarios (i.e. each row of the matrix is a sample vector). Since irradiance is very close to zero during night hours, we only consider the irradiance parameters from 5am.

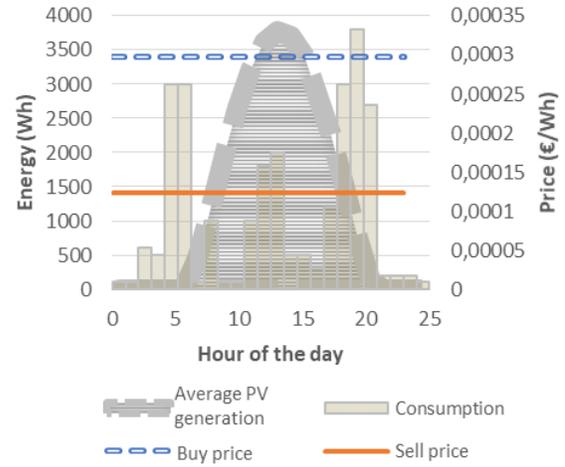

Figure 4. Power and prices profiles for each hour of the day

7pm. This will make the number of parameters describing irradiance equal to 15. With the battery capacity, the total number of inputs for our sensitivity analysis will equal 16. If we choose to set N equal to 1000 and the number of model inputs d is 16, we need to run the model on a 34000 sample matrix. Then, we can estimate the sensitivity indices based on the outputs of all run scenarios.

Results are shown in Tables I and II for first order and second order indices, respectively. Indices are highlighted with more intensified color when the value of the index is higher. The values in Table I are very similar to those of first order indices. This can be checked since first order indices sum to 1 for each row of the table. This observation suggests that there is little interaction between variables in this case. In Table I, in addition to testing for the different capacity classes, we also test cases where we shift left or right the generation curve for some hours with respect to Figure 4 (a.k.a. "No change" case) in order to have an earlier or later generation peak respectively. Table II only shows results for the "No change" case and illustrates the effect of pure interaction between any pair of input variables.

Looking at first order indices and for the different classes of battery capacities, we notice a change in the impact of battery capacity variable. Indeed, if we compute the positive netload at the end of the day (since we know the optimization strategy in this case), this will give us the battery capacity required and used to the fullest. These values are for each of the shifting cases:
- Left shift of 2: 17479.5 Wh
- Left shift of 1: 15479.5 Wh
- No change: 12932.9 Wh
- Right shift of 1: 9142.7 Wh
- Right shift of 2: 5346.8 Wh

We can see that, depending on the capacity range and how it compares to previously presented values, the impact of the capacity value changes drastically. This can be explained by the usage of the whole capacity when capacity is lower than the needed usage of a specific case: capacity variation has an important effect on the total cost since it reflects how much energy can be actually stored.

TABLE I. TOTAL ORDER SENSITIVITY INDICES

| Changes in average PV generation (in hours) | Capacity range (Wh) | 5am | 6am | 7am | 8am | 9am | 10am | 11am | 12pm | 1pm | 2pm | 3pm | 4pm | 5pm | 6pm | 7pm | capacity |
|---|---|---|---|---|---|---|---|---|---|---|---|---|---|---|---|---|---|
| Left shift of 2 | 5000 to 8000 | 0,0361 | 0,0172 | 0,0724 | 0,0434 | 0,0548 | 0,0595 | 0,0591 | 0,0488 | 0,0341 | 0,0228 | 0,0097 | 0,0147 | 0,0005 | 0,0000 | 0,0000 | 0,5435 |
| Left shift of 2 | 9000 to 12000 | 0,0277 | 0,0725 | 0,1332 | 0,0334 | 0,0422 | 0,0458 | 0,0455 | 0,0376 | 0,0263 | 0,0175 | 0,0075 | 0,0113 | 0,0004 | 0,0000 | 0,0000 | 0,5129 |
| Left shift of 2 | 15000 to 20000 | 0,0278 | 0,0730 | 0,1337 | 0,1955 | 0,1875 | 0,0460 | 0,0456 | 0,0377 | 0,0263 | 0,0176 | 0,0075 | 0,0114 | 0,0004 | 0,0000 | 0,0000 | 0,2126 |
| Left shift of 2 | 30000 to 40000 | 0,0321 | 0,0842 | 0,1542 | 0,2263 | 0,2857 | 0,0530 | 0,0526 | 0,0435 | 0,0304 | 0,0203 | 0,0087 | 0,0131 | 0,0004 | 0,0000 | 0,0000 | 0,0000 |
| Left shift of 1 | 5000 to 8000 | 0,0048 | 0,0303 | 0,0823 | 0,0258 | 0,0379 | 0,0477 | 0,0539 | 0,0496 | 0,0429 | 0,0313 | 0,0188 | 0,0085 | 0,0131 | 0,0004 | 0,0000 | 0,5644 |
| Left shift of 1 | 9000 to 12000 | 0,0049 | 0,0319 | 0,0839 | 0,0472 | 0,0386 | 0,0486 | 0,0549 | 0,0506 | 0,0437 | 0,0319 | 0,0192 | 0,0087 | 0,0134 | 0,0004 | 0,0000 | 0,5381 |
| Left shift of 1 | 15000 to 20000 | 0,0063 | 0,0416 | 0,1095 | 0,2008 | 0,2934 | 0,0634 | 0,0716 | 0,0660 | 0,0571 | 0,0416 | 0,0250 | 0,0113 | 0,0174 | 0,0005 | 0,0000 | 0,0024 |
| Left shift of 1 | 30000 to 40000 | 0,0063 | 0,0416 | 0,1094 | 0,2006 | 0,2945 | 0,0633 | 0,0716 | 0,0660 | 0,0570 | 0,0416 | 0,0250 | 0,0113 | 0,0174 | 0,0005 | 0,0000 | 0,0000 |
| No change | 5000 to 8000 | 0,0000 | 0,0047 | 0,0311 | 0,0149 | 0,0257 | 0,0376 | 0,0492 | 0,0516 | 0,0498 | 0,0449 | 0,0295 | 0,0188 | 0,0506 | 0,0128 | 0,0004 | 0,5862 |
| No change | 9000 to 12000 | 0,0000 | 0,0053 | 0,0350 | 0,0920 | 0,0521 | 0,1155 | 0,0554 | 0,0581 | 0,0560 | 0,0505 | 0,0332 | 0,0211 | 0,0569 | 0,0144 | 0,0004 | 0,3867 |
| No change | 15000 to 20000 | 0,0000 | 0,0060 | 0,0394 | 0,1037 | 0,1903 | 0,2783 | 0,0623 | 0,0654 | 0,0630 | 0,0569 | 0,0373 | 0,0238 | 0,0641 | 0,0162 | 0,0005 | 0,0000 |
| No change | 30000 to 40000 | 0,0000 | 0,0060 | 0,0394 | 0,1037 | 0,1903 | 0,2783 | 0,0623 | 0,0654 | 0,0630 | 0,0569 | 0,0373 | 0,0238 | 0,0641 | 0,0162 | 0,0005 | 0,0000 |
| Right shift of 1 | 5000 to 8000 | 0,0000 | 0,0000 | 0,0048 | 0,0306 | 0,0152 | 0,0651 | 0,0396 | 0,0481 | 0,0528 | 0,0531 | 0,0431 | 0,0300 | 0,1137 | 0,0502 | 0,0129 | 0,4543 |
| Right shift of 1 | 9000 to 12000 | 0,0000 | 0,0000 | 0,0067 | 0,0443 | 0,1162 | 0,2130 | 0,0607 | 0,0674 | 0,0740 | 0,0745 | 0,0604 | 0,0420 | 0,1595 | 0,0703 | 0,0180 | 0,0001 |
| Right shift of 1 | 15000 to 20000 | 0,0000 | 0,0000 | 0,0067 | 0,0443 | 0,1165 | 0,2129 | 0,0603 | 0,0674 | 0,0739 | 0,0744 | 0,0604 | 0,0420 | 0,1594 | 0,0703 | 0,0180 | 0,0000 |
| Right shift of 1 | 30000 to 40000 | 0,0000 | 0,0000 | 0,0067 | 0,0443 | 0,1165 | 0,2129 | 0,0603 | 0,0674 | 0,0739 | 0,0744 | 0,0604 | 0,0420 | 0,1594 | 0,0703 | 0,0180 | 0,0000 |
| Right shift of 2 | 5000 to 8000 | 0,0000 | 0,0000 | 0,0000 | 0,0057 | 0,0368 | 0,0994 | 0,1682 | 0,0457 | 0,0581 | 0,0666 | 0,0603 | 0,0519 | 0,2149 | 0,1334 | 0,0598 | 0,0028 |
| Right shift of 2 | 9000 to 12000 | 0,0000 | 0,0000 | 0,0000 | 0,0058 | 0,0381 | 0,0999 | 0,1697 | 0,0460 | 0,0584 | 0,0670 | 0,0606 | 0,0522 | 0,2160 | 0,1341 | 0,0601 | 0,0000 |
| Right shift of 2 | 15000 to 20000 | 0,0000 | 0,0000 | 0,0000 | 0,0058 | 0,0381 | 0,0999 | 0,1697 | 0,0460 | 0,0584 | 0,0670 | 0,0606 | 0,0522 | 0,2160 | 0,1341 | 0,0601 | 0,0000 |
| Right shift of 2 | 30000 to 40000 | 0,0000 | 0,0000 | 0,0000 | 0,0058 | 0,0381 | 0,0999 | 0,1697 | 0,0460 | 0,0584 | 0,0670 | 0,0606 | 0,0522 | 0,2160 | 0,1341 | 0,0601 | 0,0000 |

TABLE II. SECOND ORDER SENSITIVITY INDICES

| | 5am | 6am | 7am | 8am | 9am | 10am | 11am | 12pm | 1pm | 2pm | 3pm | 4pm | 5pm | 6pm | 7pm | capacity |
|---|---|---|---|---|---|---|---|---|---|---|---|---|---|---|---|---|
| 5am | 0,0000 | 0,0000 | 0,0000 | 0,0000 | 0,0000 | 0,0000 | 0,0000 | 0,0000 | 0,0000 | 0,0000 | 0,0000 | 0,0000 | 0,0000 | 0,0000 | 0,0000 | 0,0000 |
| 6am | 0,0000 | 0,0000 | 0,0015 | 0,0012 | 0,0017 | 0,0014 | 0,0020 | 0,0020 | 0,0015 | 0,0014 | 0,0015 | 0,0014 | 0,0014 | 0,0015 | 0,0014 | 0,0010 |
| 7am | 0,0000 | 0,0015 | 0,0000 | -0,0043 | -0,0040 | -0,0039 | -0,0043 | -0,0046 | -0,0044 | -0,0046 | -0,0042 | -0,0044 | -0,0042 | -0,0044 | -0,0041 | -0,0004 |
| 8am | 0,0000 | 0,0012 | -0,0043 | 0,0000 | 0,0009 | 0,0013 | 0,0007 | 0,0017 | 0,0017 | 0,0014 | 0,0013 | 0,0010 | 0,0008 | 0,0011 | 0,0012 | 0,0010 |
| 9am | 0,0000 | 0,0017 | -0,0040 | 0,0009 | 0,0000 | 0,0028 | 0,0022 | 0,0026 | 0,0028 | 0,0028 | 0,0029 | 0,0023 | 0,0023 | 0,0032 | 0,0025 | 0,0028 |
| 10am | 0,0000 | 0,0014 | -0,0039 | 0,0013 | 0,0028 | 0,0000 | -0,0001 | -0,0001 | 0,0001 | -0,0003 | -0,0001 | 0,0002 | -0,0001 | 0,0000 | -0,0001 | -0,0004 |
| 11am | 0,0000 | 0,0020 | -0,0043 | 0,0007 | 0,0022 | -0,0001 | 0,0000 | 0,0013 | 0,0018 | 0,0008 | 0,0021 | 0,0014 | 0,0014 | 0,0014 | 0,0012 | 0,0037 |
| 12pm | 0,0000 | 0,0020 | -0,0046 | 0,0017 | 0,0026 | -0,0001 | 0,0013 | 0,0000 | 0,0019 | 0,0032 | 0,0016 | 0,0026 | 0,0026 | 0,0011 | 0,0024 | 0,0048 |
| 1pm | 0,0000 | 0,0015 | -0,0044 | 0,0017 | 0,0028 | 0,0001 | 0,0018 | 0,0019 | 0,0000 | -0,0034 | -0,0037 | -0,0037 | -0,0058 | -0,0040 | -0,0038 | -0,0038 |
| 2pm | 0,0000 | 0,0014 | -0,0046 | 0,0014 | 0,0028 | -0,0003 | 0,0008 | 0,0032 | -0,0034 | 0,0000 | -0,0013 | -0,0009 | -0,0012 | -0,0011 | -0,0013 | -0,0018 |
| 3pm | 0,0000 | 0,0015 | -0,0042 | 0,0013 | 0,0029 | -0,0001 | 0,0021 | 0,0016 | -0,0037 | -0,0013 | 0,0000 | -0,0025 | -0,0044 | -0,0026 | -0,0026 | -0,0033 |
| 4pm | 0,0000 | 0,0014 | -0,0044 | 0,0010 | 0,0023 | 0,0002 | 0,0014 | 0,0026 | -0,0037 | -0,0009 | -0,0025 | 0,0000 | 0,0001 | 0,0004 | 0,0003 | 0,0003 |
| 5pm | 0,0000 | 0,0014 | -0,0042 | 0,0008 | 0,0023 | -0,0001 | 0,0014 | 0,0026 | -0,0058 | -0,0012 | -0,0044 | 0,0001 | 0,0000 | -0,0090 | -0,0090 | -0,0160 |
| 6pm | 0,0000 | 0,0015 | -0,0044 | 0,0011 | 0,0032 | 0,0000 | 0,0014 | 0,0011 | -0,0040 | -0,0011 | -0,0026 | 0,0004 | -0,0090 | 0,0000 | 0,0000 | 0,0010 |
| 7pm | 0,0000 | 0,0014 | -0,0041 | 0,0012 | 0,0025 | -0,0001 | 0,0012 | 0,0024 | -0,0038 | -0,0013 | -0,0026 | 0,0003 | -0,0090 | 0,0000 | 0,0000 | -0,0001 |
| capacity | 0,0000 | 0,0010 | -0,0004 | 0,0010 | 0,0028 | -0,0004 | 0,0037 | 0,0048 | -0,0038 | -0,0018 | -0,0033 | 0,0003 | -0,0160 | 0,0010 | -0,0001 | 0,0000 |

Capacity range 5000 to 8000 W

| | 5am | 6am | 7am | 8am | 9am | 10am | 11am | 12pm | 1pm | 2pm | 3pm | 4pm | 5pm | 6pm | 7pm | capacity |
|---|---|---|---|---|---|---|---|---|---|---|---|---|---|---|---|---|
| 5am | 0,0000 | 0,0000 | 0,0000 | 0,0000 | 0,0000 | 0,0000 | 0,0000 | 0,0000 | 0,0000 | 0,0000 | 0,0000 | 0,0000 | 0,0000 | 0,0000 | 0,0000 | 0,0000 |
| 6am | 0,0000 | 0,0000 | 0,0022 | 0,0017 | 0,0024 | 0,0019 | 0,0027 | 0,0027 | 0,0022 | 0,0020 | 0,0021 | 0,0021 | 0,0021 | 0,0021 | 0,0021 | 0,0011 |
| 7am | 0,0000 | 0,0022 | 0,0000 | -0,0030 | -0,0021 | -0,0022 | -0,0026 | -0,0030 | -0,0028 | -0,0029 | -0,0025 | -0,0027 | -0,0025 | -0,0028 | -0,0024 | 0,0009 |
| 8am | 0,0000 | 0,0017 | -0,0030 | 0,0000 | 0,0047 | 0,0059 | 0,0037 | 0,0060 | 0,0067 | 0,0062 | 0,0055 | 0,0052 | 0,0044 | 0,0051 | 0,0055 | 0,0058 |
| 9am | 0,0000 | 0,0024 | -0,0021 | 0,0047 | 0,0000 | 0,0049 | 0,0052 | 0,0051 | 0,0055 | 0,0054 | 0,0058 | 0,0045 | 0,0047 | 0,0060 | 0,0051 | 0,0153 |
| 10am | 0,0000 | 0,0019 | -0,0022 | 0,0059 | 0,0049 | 0,0000 | -0,0041 | -0,0044 | -0,0037 | -0,0045 | -0,0033 | -0,0031 | -0,0029 | -0,0036 | -0,0038 | 0,0108 |
| 11am | 0,0000 | 0,0027 | -0,0026 | 0,0037 | 0,0052 | -0,0041 | 0,0000 | 0,0007 | 0,0012 | 0,0001 | 0,0015 | 0,0007 | 0,0008 | 0,0008 | 0,0006 | 0,0030 |
| 12pm | 0,0000 | 0,0027 | -0,0030 | 0,0060 | 0,0051 | -0,0044 | 0,0007 | 0,0000 | 0,0012 | 0,0026 | 0,0009 | 0,0019 | 0,0020 | 0,0003 | 0,0017 | 0,0048 |
| 1pm | 0,0000 | 0,0022 | -0,0028 | 0,0067 | 0,0055 | -0,0037 | 0,0012 | 0,0012 | 0,0000 | -0,0060 | -0,0063 | -0,0064 | -0,0088 | -0,0067 | -0,0065 | -0,0061 |
| 2pm | 0,0000 | 0,0020 | -0,0029 | 0,0062 | 0,0054 | -0,0045 | 0,0001 | 0,0026 | -0,0060 | 0,0000 | -0,0008 | -0,0004 | -0,0006 | -0,0005 | -0,0008 | -0,0005 |
| 3pm | 0,0000 | 0,0021 | -0,0025 | 0,0055 | 0,0058 | -0,0033 | 0,0015 | 0,0009 | -0,0063 | -0,0008 | 0,0000 | -0,0049 | -0,0071 | -0,0050 | -0,0050 | -0,0025 |
| 4pm | 0,0000 | 0,0021 | -0,0027 | 0,0052 | 0,0045 | -0,0031 | 0,0007 | 0,0019 | -0,0064 | -0,0004 | -0,0049 | 0,0000 | -0,0003 | 0,0001 | 0,0000 | -0,0003 |
| 5pm | 0,0000 | 0,0021 | -0,0025 | 0,0044 | 0,0047 | -0,0029 | 0,0008 | 0,0020 | -0,0088 | -0,0006 | -0,0071 | -0,0003 | 0,0000 | -0,0066 | -0,0066 | -0,0129 |
| 6pm | 0,0000 | 0,0021 | -0,0028 | 0,0051 | 0,0060 | -0,0036 | 0,0008 | 0,0003 | -0,0067 | -0,0005 | -0,0050 | 0,0001 | -0,0066 | 0,0000 | -0,0017 | -0,0016 |
| 7pm | 0,0000 | 0,0021 | -0,0024 | 0,0055 | 0,0051 | -0,0038 | 0,0006 | 0,0017 | -0,0065 | -0,0008 | -0,0050 | 0,0000 | -0,0066 | -0,0017 | 0,0000 | 0,0001 |
| capacity | 0,0000 | 0,0011 | 0,0009 | 0,0058 | 0,0153 | 0,0108 | 0,0030 | 0,0048 | -0,0061 | -0,0005 | -0,0025 | -0,0003 | -0,0129 | -0,0016 | 0,0001 | 0,0000 |

Capacity range 9000 to 12000 W

| | 5am | 6am | 7am | 8am | 9am | 10am | 11am | 12pm | 1pm | 2pm | 3pm | 4pm | 5pm | 6pm | 7pm | capacity |
|---|---|---|---|---|---|---|---|---|---|---|---|---|---|---|---|---|
| 5am | 0,0000 | 0,0000 | 0,0000 | 0,0000 | 0,0000 | 0,0000 | 0,0000 | 0,0000 | 0,0000 | 0,0000 | 0,0000 | 0,0000 | 0,0000 | 0,0000 | 0,0000 | 0,0000 |
| 6am | 0,0000 | 0,0000 | 0,0015 | 0,0009 | 0,0022 | 0,0013 | 0,0021 | 0,0021 | 0,0015 | 0,0013 | 0,0015 | 0,0014 | 0,0014 | 0,0014 | 0,0014 | 0,0014 |
| 7am | 0,0000 | 0,0015 | 0,0000 | 0,0012 | 0,0023 | 0,0025 | 0,0016 | 0,0012 | 0,0014 | 0,0012 | 0,0017 | 0,0015 | 0,0017 | 0,0014 | 0,0018 | 0,0017 |
| 8am | 0,0000 | 0,0009 | 0,0012 | 0,0000 | 0,0006 | 0,0021 | 0,0003 | 0,0027 | 0,0038 | 0,0030 | 0,0023 | 0,0019 | 0,0011 | 0,0018 | 0,0023 | 0,0019 |
| 9am | 0,0000 | 0,0022 | 0,0023 | 0,0006 | 0,0000 | 0,0066 | 0,0042 | 0,0054 | 0,0059 | 0,0060 | 0,0063 | 0,0045 | 0,0046 | 0,0073 | 0,0051 | 0,0055 |
| 10am | 0,0000 | 0,0013 | 0,0025 | 0,0021 | 0,0066 | 0,0000 | 0,0008 | 0,0007 | 0,0011 | 0,0001 | 0,0007 | 0,0016 | 0,0006 | 0,0009 | 0,0006 | 0,0006 |
| 11am | 0,0000 | 0,0021 | 0,0016 | 0,0003 | 0,0042 | 0,0008 | 0,0000 | -0,0010 | -0,0004 | -0,0017 | -0,0001 | -0,0010 | -0,0009 | -0,0009 | -0,0011 | -0,0010 |
| 12pm | 0,0000 | 0,0021 | 0,0012 | 0,0027 | 0,0054 | 0,0007 | -0,0010 | 0,0000 | 0,0050 | 0,0066 | 0,0047 | 0,0059 | 0,0060 | 0,0040 | 0,0056 | 0,0057 |
| 1pm | 0,0000 | 0,0015 | 0,0014 | 0,0038 | 0,0059 | 0,0011 | -0,0004 | 0,0050 | 0,0000 | -0,0047 | -0,0051 | -0,0052 | -0,0079 | -0,0055 | -0,0054 | -0,0054 |
| 2pm | 0,0000 | 0,0013 | 0,0012 | 0,0030 | 0,0060 | 0,0001 | -0,0017 | 0,0066 | -0,0047 | 0,0000 | -0,0016 | -0,0011 | -0,0014 | -0,0013 | -0,0015 | -0,0015 |
| 3pm | 0,0000 | 0,0015 | 0,0017 | 0,0023 | 0,0063 | 0,0007 | -0,0001 | 0,0047 | -0,0051 | -0,0016 | 0,0000 | -0,0023 | -0,0047 | -0,0024 | -0,0024 | -0,0024 |
| 4pm | 0,0000 | 0,0014 | 0,0015 | 0,0019 | 0,0045 | 0,0016 | -0,0010 | 0,0059 | -0,0052 | -0,0011 | -0,0023 | 0,0000 | 0,0002 | 0,0005 | 0,0004 | 0,0004 |
| 5pm | 0,0000 | 0,0014 | 0,0017 | 0,0011 | 0,0046 | 0,0006 | -0,0009 | 0,0060 | -0,0079 | -0,0014 | -0,0047 | 0,0002 | 0,0000 | 0,0009 | 0,0009 | 0,0009 |
| 6pm | 0,0000 | 0,0014 | 0,0014 | 0,0018 | 0,0073 | 0,0009 | -0,0009 | 0,0040 | -0,0055 | -0,0013 | -0,0024 | 0,0005 | 0,0009 | 0,0000 | -0,0035 | -0,0035 |
| 7pm | 0,0000 | 0,0014 | 0,0018 | 0,0023 | 0,0051 | 0,0006 | -0,0011 | 0,0056 | -0,0054 | -0,0015 | -0,0024 | 0,0004 | 0,0009 | -0,0035 | 0,0000 | 0,0003 |
| capacity | 0,0000 | 0,0014 | 0,0017 | 0,0019 | 0,0055 | 0,0006 | -0,0010 | 0,0057 | -0,0054 | -0,0015 | -0,0024 | 0,0004 | 0,0009 | -0,0035 | 0,0003 | 0,0000 |

Capacity range 15000 to 20000 W

When the capacity class becomes higher, the index will drop to zero. Hence, in this study, the sensitivity analysis can allow us to identify the "best" capacity class with respect to our energy management system. Indeed, the second class batteries seem to best fit storage needs based on the "No change" test case.

When battery capacity has a significant sensitivity index, the earliest hours with the highest netload value are the most impacting (i.e., hours 7am-9am). This suggests that day-ahead planning of charging is taking place during these hours.

Looking at the hour for irradiance, we can see that the sensitivity index is proportional to the irradiance value. At 10am and 11am, the prediction error has the highest impact on the estimated total cost and optimization results.

For second order indices in Table II, we can observe a change in results depending on the capacity class. The first class of capacity presents a high sensitivity to irradiance at 9pm, 11am and 12pm coupled with the subsequent hours of each as well as capacity. 9am represents the hour at which the battery will be fully charged. So, the state will be impacted in subsequent hours with respect to discharge possibilities. 11am and 12pm are hours during which the netload is very low. Then, variability of generation affects cost significantly.

The second class of capacity has a sensitivity that is the most significant for hours 8am and 9am coupled with variables representing subsequent hours and capacity. This is unexpected since usually hours at which the battery is fully charged are the most impacting in combination with subsequent hours and the capacity (1pm for this capacity class). However, 8am and 9am are crucial hours for charging the battery and can affect decisions taken subsequently.

For large batteries (the third capacity class), the most important sensitivity from coupling variables is that of hours 9am and 12pm. In this case, 9am presents a peak in netload and 12pm represents the hour. In this study at which the battery might become the fullest (since charging is constrained by energy consumption in subsequent hours).

As a final observation, we can notice that second order sensitivity indices become more significant when the capacity is larger.

## IV. CONCLUSION

In this paper, we analyze the sensitivity of radiation forecast on the total cost incurred in a home equipped with a flexible device and an energy management system. We use a black-box technique to conduct the sensitivity analysis. The chosen technique takes into account requirements of minimizing the time and computation required through quasi-random sampling. For the proposed energy management system based on day-ahead scheduling, the analysis results reflect well the charging strategy. The analysis also allows us to validate the use (or not) of the full battery capacity. Although we consider a simple scenario, our system allows to analyze more complex energy management systems and can integrate their interaction with different pricing schemes. As a perspective, it is possible to study the impact of prediction error on a city. A decentralized and distributed solution can reduce the complexity of the sensitivity study such as proposed in [14] [15] to regularize the mismatches of supply and demand.

ACKNOWLEDGMENT

This research work has been carried out under the leadership of the Institute of Technological Research SystemX, within the scope of a joint collaboration with Sherpa Engineering (France) and Reuniwatt (France) in the Smart City Energy Analytics project.

REFERENCES

[1] M. D. McKay, "Evaluating prediction uncertainty, Technical Report NUREG/CR-6311," U.S. Nuclear Regulatory Commission and Los Alamos National Laboratory, 1995.

[2] H. Frey, A. Mokhtari, and T. Danish, "Evaluation of selected sensitivity Analysys methods Based Upon Applications to two Food safety Process risk models," Reigh-North Carolina, North Carolina State University, 2003.

[3] P. Heiselberg et al., "Application in sensitivity Analysis in design of sustainable buildings," Renewable Energy, pp. 2030-2036, 2009.

[4] A.-T. Nguyen and S. Reiter, "A performance comparison of sensitivity analysis methods for building energy models," Research Article Building Thermal, Lighting, and Acoustics Modeling, pp. 651-664, 2015.

[5] I. M. Sobol, "Sensitivity estimates for nonlinear mathematical models," MMCE 1, pp. 407-414, 1993.

[6] I. M. Sobol, "Global sensitivity indices for nonlinear mathematical models and their Monte Carlo estimates.," Mathematics and Computers in Simulation, pp. 271-280, 2001.

[7] R. I. Cukier, C. M. Fortuin, K. E. Shuler, A. G. Petschek, and J. H. Schaibly, "Study of the sensitivity of coupled reaction systems to uncertainties in rate coefficients I Theory," The Journal of Chemical Physics 59, pp. 3873-3878, 2001.

[8] Phisim library. [retrieved: April, 2019]. from https://www.sherpa-eng.com/produits/phisim/.

[9] A. Saltelli, K. Chan, and E. M. Scott, "Sensitivity Analysis," Chichester : Wiley Series in Probability and Statistics, 2000.

[10] M. Ratto, S. Tarantola, and A. Saltelli, "Estimation of importance indicators for correlated inputs," ESREL2001, 2001.

[11] A. Saltelli, "Making best use of model evaluations to compute sensitivity indices," Computer Physics Communications, pp. 280-297, 2002.

[12] A. Saltelli et al., "Variance based sensitivity analysis of model output. Design and estimator for the total sensitivity index," Computer Physics Communications, pp. 259-270, 2010.

[13] A. Henry, M. Sinquin, and C. Boudesocque, "Prospective d'évolution de la consommation électrique domestique à l'horizon 2030," INP Toulouse, ENSEEIHT, ENSIACET, 2011.

[14] Pournaras, E., Pilgerstorfer, P. and Asikis, T., 2018. Decentralized collective learning for self-managed sharing economies. ACM Transactions on Autonomous and Adaptive Systems (TAAS), 13(2), p.10.

[15] Pournaras, E., Yao, M. and Helbing, D., 2017. Self-regulating supply–demand systems. Future Generation Computer Systems, 76, pp.73-91.